\documentclass[epj]{webofc}
\usepackage[varg]{txfonts}   
\usepackage{graphicx}
\usepackage{amssymb}
\usepackage{amsmath}

\wocname{Proceedings of ``Wigner 111 -- Colourful \& Deep''}
\begin{document}
\title{Colored condensates deep inside neutron stars}
%
%

\author{%
David Blaschke\inst{1,2}
}
\institute{Institute for Theoretical Physics, University of Wroclaw (Poland)
\and 
Bogoliubov Laboratory for Theoretical Physics, JINR Dubna (Russia)}
\abstract{%
It is demonstrated how in the absence of solutions for QCD under conditions 
deep inside compact stars an equation of state can be obtained within a model 
that is built on the basic symmetries of the QCD Lagrangian, in particular
chiral symmetry and color symmetry.
While in the vacuum the chiral symmetry is spontaneously broken, it gets 
restored at high densities. 
Color symmetry, however, gets broken simultaneously
by the formation of colorful diquark condensates. 
It is shown that a strong diquark condensate in cold dense quark matter is 
essential for supporting the possibility that such states could exist in the
recently observed pulsars with masses of 2 $M_\odot$. 
}
\maketitle
\section{Color superconductivity in the QCD phase diagram}

Since Eugene Wigner was awarded the Nobel prize for his discovery and 
application of fundamental symmetry principles, particularly in 
nuclear and elementary particle physics, our understanding of strong 
interactions within quantum chromodynamics (QCD) is based more than ever 
on these concepts. 
I will elucidate this statement on the example of the QCD phase diagram and
the question for quark matter in compact stars. 
Progress has been made for understanding the mass spectrum of hadrons and 
also their interactions on the basis of ab-inito simulations using the QCD 
Lagrangian on the lattice. 
Extensions of such calculations to finite temperatures have provided 
detailed numerical data on the QCD phase transition but are limited to the 
domain of vanishing or low baryon densities, not applicable for the whole QCD 
phase diagram and in particular not for compact star interiors, see 
\cite{Karsch:2013naa} for a recent overview. 
In this situation, effective models for low-energy QCD are investigated 
instead which share basic symmetries with the QCD Lagrangian and allow to
study the patterns of their dynamical breaking and restoration under extreme
conditions of temperature and baryon density (or chemical potential), 
spanning the basic axes of the QCD phase diagram, see Fig.~1. 
The workhorse of this kind of studies is the Nambu--Jona-Lasinio (NJL) model
\cite{Klevansky:1992qe,Hatsuda:1994pi,Buballa:2003qv},
recently extended by the coupling to a background gluon field in the Polyakov
gauge with a meanfield Potential ${\cal U}$ \cite{Fukushima:2013rx},
\begin{equation}
 {\cal L} = 
\bar{q}(-i\gamma_\mu D_\mu + \hat{m} + \gamma_0\hat{\mu})q
+ {\cal L}_{\rm int} + {\cal U}(T, \Phi)~,
\end{equation}
representing the (approximate) chiral symmetry in the free Dirac 
fermion Lagrangian and the 4-fermion interaction model
in the scalar-pseudoscalar sector 
\begin{eqnarray}
{\cal{L}}_{\rm int} = 
 G_S\left\{\sum_{a=0}^{8}\left[(\bar{q}\tau_a q)^2
+(\bar{q}\tau_ai\gamma_5 q)^2 \right] 
+ \eta_V(\bar{q}i\gamma_0q)^2
+ \eta_D\sum_{a=2,5,7}
(\bar{q}i\gamma_5 \tau_a \lambda_a C \bar{q}^T)(q^TCi\gamma_5\tau_a\lambda_a q)
\right\}~,
\end{eqnarray}
augmented here with 4-fermion couplings in the isoscalar vector  and 
scalar diquark  interaction channels
with the dimensionless coupling strengths $\eta_V$ and $\eta_D$, resp., 
which are free parameters of the model, see \cite{Contrera:2012wj} for 
details.
While  $\eta_D$ to determines the onset of the quark matter phase, $\eta_V$ 
is essential for its stiffness. 
The thermodynamic potential of this model in the meanfield approximation can 
be given the closed form 
\begin{eqnarray}
\Omega\left(T,\{\mu\}\right) &=& \frac{\phi_u^2 + \phi_d^2 + \phi^2_s}{8 G_S} 
- \frac{\omega^2_u + \omega^2_d + \omega^2_s}{8 G_V} 
+ \frac{\Delta^2_{ud} + \Delta^2_{us} + \Delta^2_{ds}}{4G_D} \nonumber\\
&-& \int{\frac{d^3p}{(2\pi)^3}}\sum_{n=1}^{18}\left[E_n + 
2T\ln\left(1 + e^{-E_n/T}\right)\right]
+ \Omega_{lep} - \Omega_0 + {\cal U}(\Phi,T)~,
\end{eqnarray}
where for compact star applications the lepton contribution has been added 
and a vacuum contribution ($\Omega_0$) has been subtracted
\cite{Blaschke:2005uj}. 
The meanfields $\phi_f$, $\Delta_{ik}$ and 
$\Phi=(1/N_c){\rm Tr}_c \exp[i\beta(\phi_3\lambda_3+\phi_8\lambda_8)]$
are the order parameters signalling chiral symmetry breaking ($\chi$SB), 
color superconductivity (2SC and CFL) and deconfinement, respectively.
Their values are determined by the necessary condition for a minimum of the 
thermodynamic potential (gap equations) under the constraint of vanishing 
color charge denity (the last two equations)
\begin{equation}
\frac{\partial\Omega}{\partial\phi_f}=
\frac{\partial\Omega}{\partial\Delta_{ik}}=
\frac{\partial\Omega}{\partial\phi_3}=
\frac{\partial\Omega}{\partial\phi_8}=0
~;~~
\frac{\partial\Omega}{\partial\mu_3}=
\frac{\partial\Omega}{\partial\mu_8}=0~,
\end{equation}
that need to be solved self-consistently. 
The results are shown in the phase diagram of Fig.~1 (left panel), where in 
the plane of temperature $T$ and quark chemical potential $\mu$ the basic 
QCD phases are shown as regions with nonvanishing order parameters.
An interesting artifact of the meanfield approximation is the occurrence of a
region with negative baryon density $n_B<0$ which signals a thermodynamic 
instability \cite{GomezDumm:2008sk}. 
It is cured by the occurrence of hadronic matter (beyond meanfield) in the 
region where chiral symmetry is broken and confinement prevails, see the right
panel of Fig.~1 where a quark-hadron phase transition is obtained by a Maxwell
construction between the meanfield quark matter model and a density-dependent 
relativistic meanfield model of nuclear matter.
It is interesting to note that the Polyakov-loop suppression of colored quark
excitations in the region of the phase diagram where $\Phi\ll 1$ leads to a
strong reduction of the quark Pauli blocking effect and thus to a strong 
increase of the critical temperature of the color superconducting 2SC phase.
This raises hopes that this exotic phase of QCD could be accessible 
\cite{Blaschke:2010ka} in future dedicated heavy-ion collision experiments 
like at FAIR-CBM in Darmstadt and NICA-MPD or BM@N in Dubna 
\cite{Klahn:2011au}.
 
\begin{figure}[!ht]
\centering
\includegraphics[width=0.49\textwidth]{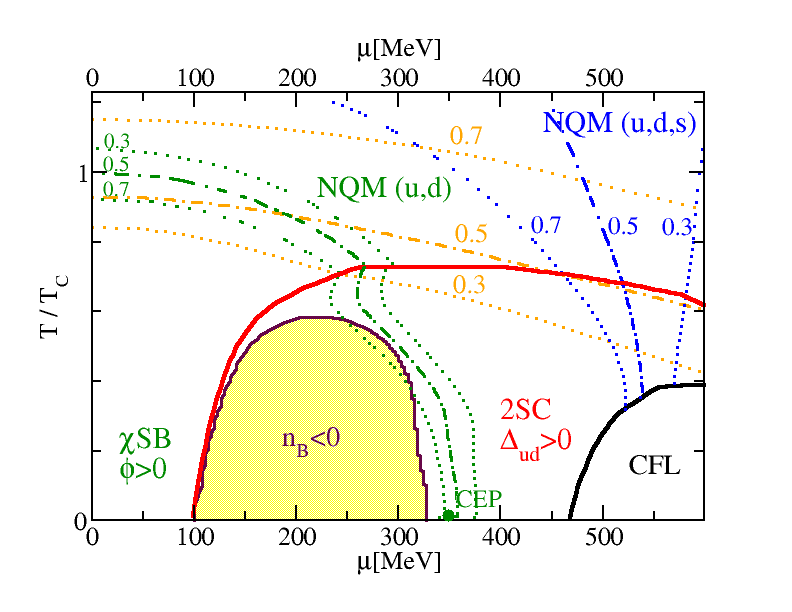}
\includegraphics[width=0.49\textwidth]{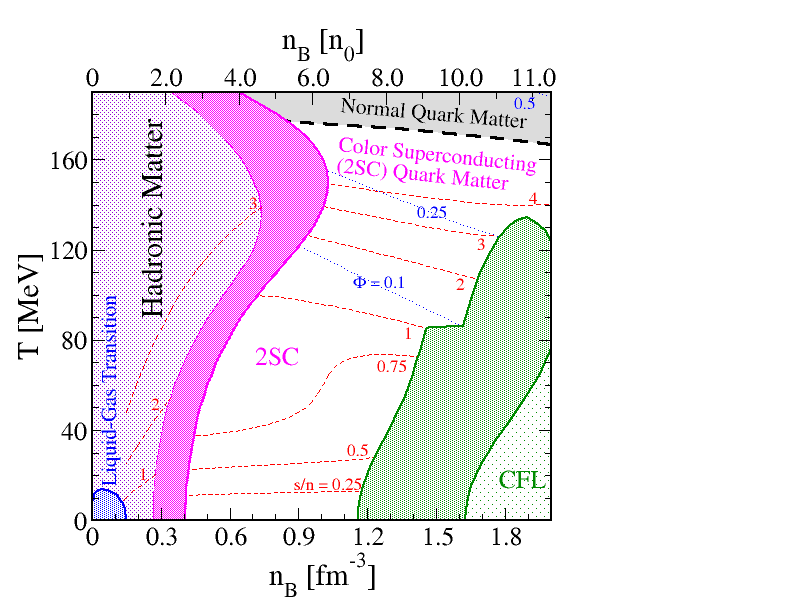}
\caption{\label{fig:PHD}
Left: Phase diagram of color superconducting 3-flavor PNJL model at meanfield
level in the $T$-$\mu$ plane with a thermodynamic instability $n_B<0$ in the 
coexistence region of 2SC and $\chi$SB phases. 
Right: Phase diagram with first order transition (Maxwell construction) to 
hadronic matter described in the DD2 
model.
}
\end{figure}

\section{Color superconducting phases in compact stars}

For the application of the above described low-energy QCD model to study the 
composition of compact star interiors one has to embody the constraints of 
electric neutrality and $\beta-$equilibrium to the thermodynamic potential
and then use the resulting $T=0$ equation of state (EoS) for the pressure
$p(\mu)=\Omega(T=0,\{\mu\})$ from which all other thermodynamic functions
can be derived which are required to solve the Tolman-Oppenheimer-Volkoff 
equations of compact star stability.
To this end, we use a hybrid EoS obtained by a Maxwell construction with the
DBHF EoS for hadronic matter.

\begin{figure}[!htb]
\centering
\includegraphics[width=0.49\columnwidth]{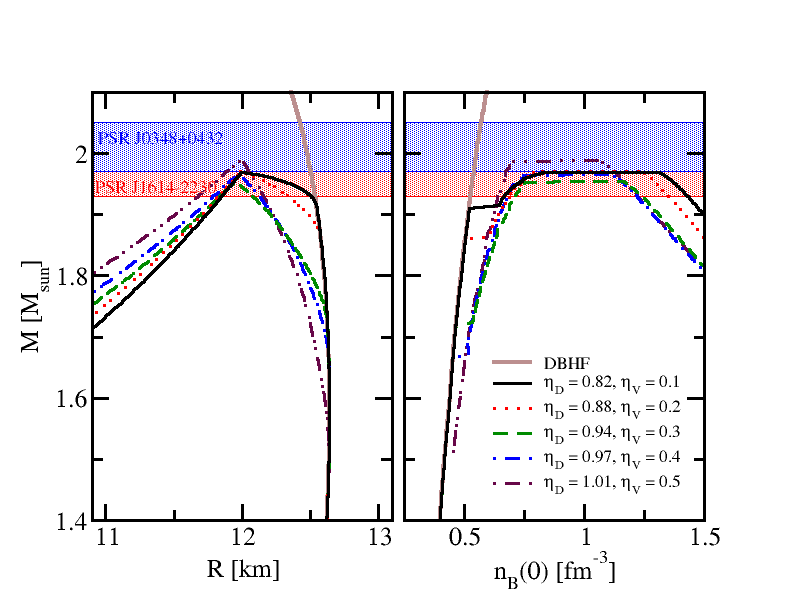}
\includegraphics[width=0.49\columnwidth]{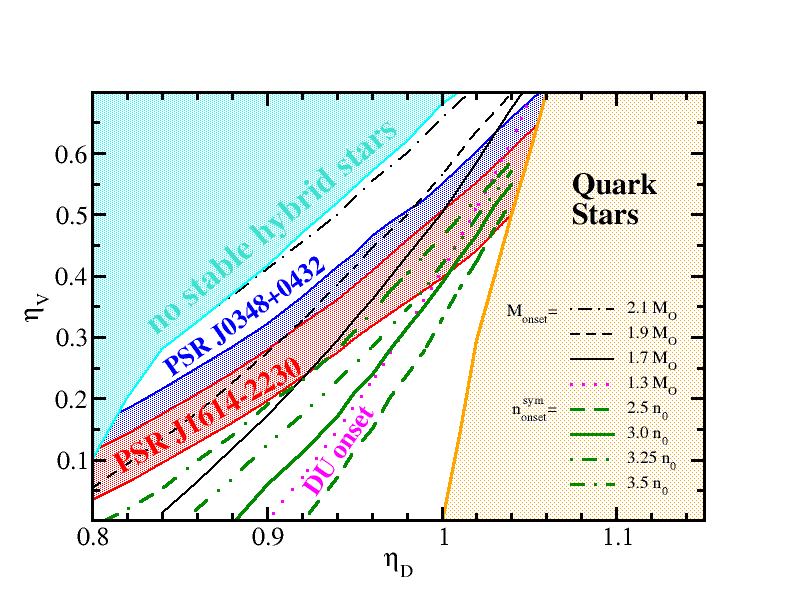}
\caption{
Left: Mass-radius and mass-central density sequences for parameter sets 
allowing hybrid stars which fulfill the 2 $M_\odot$ mass constraints.
Right: Region of parameter values for the vector meson ($\eta_V$) and scalar
diquark ($\eta_D$) coupling strengths where stable hybrid stars with quark 
matter cores are possible (between cyan and orange domains) and where the 
2 $M_\odot$ mass constraints for the pulsars PSR J0348+0432 (blue) and 
PSR J1614-2230 (red) are fulfilled. 
\label{fig:Monster2}}
\end{figure}
Results are shown in Fig.~2 where in the left panel sequences of compact star 
configurations are given in the mass ($M$) vs. radius ($R$) and mass vs. 
central baryon density $n_B(0)$ plane, resp., for given choices of the 
parameter pairs $(\eta_V,\eta_D)$ which allow to describe massive stars of 
the kind of the two recently discovered $2M_\odot$ pulsars 
\cite{Demorest:2010bx,Antoniadis:2013pzd} as hybrid stars with a color 
superconducting quark matter core.
Note that an increase of the diquark coupling entails a lowering of the 
critical density for the onset of 2SC quark matter but requires a 
simultaneous stiffening by an increase of the vector coupling in order to 
still allow a sufficiently high maximum mass in accordance with the 
measurements \cite{Demorest:2010bx,Antoniadis:2013pzd}.
A second observation is that the maximum allowable mass of a sequence is 
determined by the onset of strange quark matter formation (CFL phase) which
softens the EoS so much that all configurations with strange quark matter 
cores lie on an unstable branch.
The results of these systematic studies within the (P)NJL model for quark 
matter \cite{Klahn:2013kga} are summarized in the right panel of Fig.~2.
A conclusion drawn from this study is that should the high-mass pulsars be 
hybrid stars with a quark matter core the latter has to be in color 
superconducting phase!
Another conclusion from such a study is that the critical density for the 
deconfinement transition in heavy-ion collisions (symmetric nuclear matter) 
should exceed 3.5 saturation densities unless for the diquark coupling 
holds $\eta_V > 1.0$.
The appearance of quark matter in the cores of compact stars is a possibility 
to solve problems that occur with purely hadronic scenarios such as the direct
Urca cooling problem \cite{Klahn:2006iw}
and the hyperon puzzle \cite{Lastowiecki:2011hh}. 
The question arises: Can one detect by observations of compact stars
a deconfinement phase transition in their interior?
Recently, it has been suggested that: yes, if our two high-mass pulsars would
be so-called ``mass twins'', i.e. have very different radii so that the more
compact twin would be the one with a quark matter interior while the larger 
star would consist of hadronic matter - this would be possible at the same 
high mass only with a strong phase transition with sufficiently large latent 
heat \cite{Blaschke:2013ana}!  
Such transitions with a coexistence of two phases are known to produce 
structures like bubbles and droplets (pasta phase) of different shapes and 
sizes governed by surface tension and Coulomb effects, see  
\cite{Yasutake:2014oxa} and references therein. For the calculation of such 
pasta phases one uses the very successful concept of the Wigner-Seitz cell
\cite{Wigner:1933zz}.
A strong QCD phase transition might contribute to solving one of the 
long-standing puzzles in Astrophysics: the quest for a supernova explosion 
mechanism \cite{Fischer:2014uua}.

\subsection*{Acknowledgements}
Inspiring discussions with many colleagues and the high intellectual spirit 
of this Conference celebrating Eugene Wigners 111$^{\rm th}$ birthday are 
gratefully acknowledged.
This work was supported in part by NCN under grant no. UMO-2011/02/A/ST2/00306,
by MNiSW under grant no. 1009/S/IFT/14 and by the COST Action MP1304 
``Exploring fundamental physics with compact stars (NewCompStar)''.


\begin{thebibliography}{9}
\bibitem{Karsch:2013naa} 
  F.~Karsch,
  Acta Phys.\ Polon.\ Supp.\  {\bf 7}, no. 1, 117 (2014).

\bibitem{Klevansky:1992qe} 
  S.~P.~Klevansky,
  Rev.\ Mod.\ Phys.\  {\bf 64}, 649 (1992).

\bibitem{Hatsuda:1994pi} 
  T.~Hatsuda and T.~Kunihiro,
  Phys.\ Rept.\  {\bf 247}, 221 (1994).

\bibitem{Buballa:2003qv} 
  M.~Buballa,
  Phys.\ Rept.\  {\bf 407}, 205 (2005).


\bibitem{Fukushima:2013rx} 
  K.~Fukushima and C.~Sasaki,
  Prog.\ Part.\ Nucl.\ Phys.\  {\bf 72}, 99 (2013).

\bibitem{Contrera:2012wj} 
  G.~A.~Contrera, A.~G.~Grunfeld and D.~B.~Blaschke,
  Phys. Part. Nucl. Lett. {\bf 11}, 342 (2014).

\bibitem{Blaschke:2005uj} 
  D.~Blaschke, S.~Fredriksson, H.~Grigorian, A.~M.~\"Oztas and F.~Sandin,
  Phys.\ Rev.\ D {\bf 72}, 065020 (2005).

\bibitem{GomezDumm:2008sk} 
  D.~Gomez Dumm, D.~B.~Blaschke, A.~G.~Grunfeld and N.~N.~Scoccola,
  Phys.\ Rev.\ D {\bf 78}, 114021 (2008).

\bibitem{Blaschke:2010ka} 
  D.~B.~Blaschke, F.~Sandin, V.~V.~Skokov and S.~Typel,
  Acta Phys.\ Polon.\ Supp.\  {\bf 3}, 741 (2010).

\bibitem{Klahn:2011au} 
  T.~Kl\"ahn, D.~Blaschke and F.~Weber,
  Phys.\ Part.\ Nucl.\ Lett.\  {\bf 9}, 484 (2012).



\bibitem{Demorest:2010bx} 
  P.~Demorest {\it et al.}, 
  Nature {\bf 467}, 1081 (2010).

\bibitem{Antoniadis:2013pzd} 
  J.~Antoniadis {\it et al.}, 
  Science {\bf 340}, 6131 (2013).


\bibitem{Klahn:2013kga} 
  T.~Kl\"ahn, D.~B.~Blaschke and R.~Łastowiecki,
  Phys.\ Rev.\ D {\bf 88}, 085001 (2013).


\bibitem{Klahn:2006iw} 
  T.~Kl\"ahn et al., 
  Phys.\ Lett.\ B {\bf 654}, 170 (2007).

\bibitem{Lastowiecki:2011hh} 
  R.~Lastowiecki, D.~Blaschke, H.~Grigorian and S.~Typel,
  Acta Phys.\ Polon.\ Supp.\  {\bf 5}, 535 (2012).


\bibitem{Blaschke:2013ana} 
  D.~Blaschke, D.~E.~Alvarez-Castillo and S.~Benic,
  PoS CPOD {\bf 2013}, 063 (2013).

\bibitem{Yasutake:2014oxa} 
N.~Yasutake et al., 
  Phys.\ Rev.\ C {\bf 89}, 065803 (2014).

\bibitem{Wigner:1933zz} 
  E.~Wigner and F.~Seitz,
  Phys.\ Rev.\  {\bf 43}, 804 (1933).

\bibitem{Fischer:2014uua} 
  T.~Fischer, T.~Klähn, I.~Sagert, M.~Hempel and D.~Blaschke,
  Acta Phys.\ Polon.\ Supp.\  {\bf 7}, 153 (2014).


\end{thebibliography}
\end{document}